\documentclass[pre,twocolumn,aps,showpacs,superscriptaddress]{revtex4}
\pdfoutput=1
\usepackage{graphics}
\usepackage{epsfig}
\usepackage{amsmath}
\usepackage{amssymb}
\usepackage{bbm}
\usepackage{bm}
\usepackage{color}

\newcommand{\pf}{P(f)}

\begin{document}
\title{The force network ensemble for granular packings }

\author{Brian P.~Tighe}
\affiliation{Instituut-Lorentz, Universiteit Leiden, Postbus 9506, 2300 RA Leiden, The Netherlands}
\author{Jacco H.~Snoeijer}
\affiliation{Physics of Fluids Group, University of Twente, 7500 AE Enschede, The Netherlands}
\author{Thijs J.~H.~Vlugt}
\affiliation{Delft University of Technology, Process \& Energy Laboratory, Leeghwaterstraat 44, 2628 CA Delft, The Netherlands}
\author{Martin van Hecke}
\affiliation{Kamerlingh Onnes Lab, Universiteit Leiden, Postbus 9504, 2300 RA Leiden, The Netherlands}

\date{\today}

\begin{abstract}
For packings of hard but not perfectly rigid particles, the length scales that govern the
packing geometry and the contact forces are well separated. This
separation of length scales is  explored in the force network
ensemble, where one studies the space of allowed force
configurations for a given, frozen contact geometry. Here we review results of this approach, which yields
nontrivial predictions for the effect of packing dimension and anisotropy on the contact force distribution $P(f)$, the
response to overall shear and point forcing, all of which can be studied
in great numerical detail. Moreover, there are emerging analytical approaches
that very effectively capture, for example, the form of force distributions.
\end{abstract}

\maketitle

\begin{figure}[tbp]
\centering
\includegraphics[clip,width=1.0\linewidth]{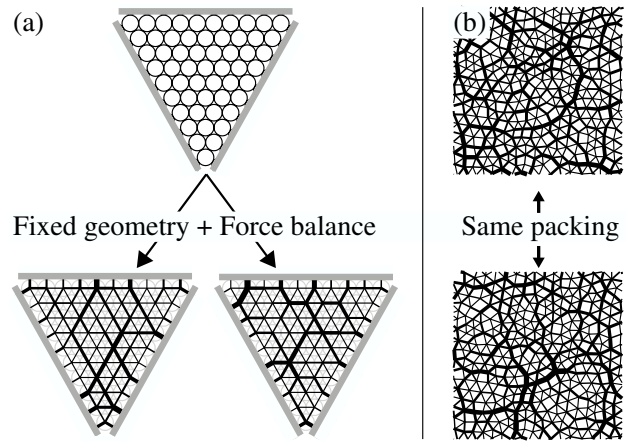}
\caption{Illustration of force networks studied in the Force Network Ensemble. Lines between disks indicate contact forces; larger forces have thicker lines. For a single ordered (a) or disordered (b) contact network, many different configurations of noncohesive contact forces, or force networks, satisfy force balance on every grain. The FNE comprises all balanced force networks on a fixed contact network. (From \cite{snoeijer04a}.) }
\label{fig:FNE}
\end{figure}

Force networks are a striking feature of granular media \cite{jaeger96,radjai96,howell99,majmudar05,vanhecke09}
--- see Fig.~\ref{fig:FNE}. This organization of the contact forces between individual grains has fascinated physicists for decades
\cite{dantu57,josselin69,janssen,janssensperl}. Why are these contact forces interesting? First, the
fluctuations of the forces appear to be unexpectedly strong, with early
measurements \cite{liu95,mueth98} and models \cite{coppersmith96} indicating that the probability distribution function of the contact forces, $P(f)$, is not narrowly distributed about its mean, but instead decays exponentially at large forces. Second, the spatial organization of the strong contact forces in so-called force networks plays an important role in the memory effects exhibited by granular media \cite{vanel99,toiya04}. Third, predicting the mechanical properties of granular assemblies is a central goal of granular theory \cite{nedderman,wittmer96,bouchaudleshouches,metzger08,liu06,depken06,henkes09}, and these stresses ultimately originate from the contact forces and their organization.

From a microscopic point of view, the contact forces arise from
deformations of the constituent grains. This perspective is useful
when developing a detailed numerical recipe, but theoretically
unwieldy. Because granular media are highly disordered the details
of the contact geometry play a crucial role, response to a load is
not affine, and macroscopic elasticity cannot be inferred directly
from a microscopic force law \cite{vanhecke09}. Additionally, grain deformations are typically very small --- a steel ball of 1 mm deforms only by a
nanometer under its own weight, and buried under a pile of beads
one meter thick, typical deformations are of order of 100 nm. This leads to a strong separation of the scale governing the
geometry of the packing and the scale governing the contact
forces. 

In this paper we review the Force Network Ensemble (FNE) \cite{snoeijer04a}, which is
a model that relies on the separation of scales relevant for the
contact forces (nanometers) and the particle scale (millimeters).
The central idea is that, for a given packing geometry, many
different microscopic configurations of noncohesive forces satisfy force (and
possibly torque) balance on each grain, while also satisfying
boundary conditions in terms of the applied stresses. In other
words, the forces can be seen as underdetermined. As a simple
example of force indeterminacy, one may think of the forces acting
on a ladder that rests against a wall under an angle --- a range
of contact forces is possible to keep the ladder in balance. This
indeterminacy carries over to collections of grains, as is
illustrated in Fig.~\ref{fig:FNE}. Both ordered and disordered
packings allow for many different force networks that respect
mechanical equilibrium on each grain.
By averaging over all possible force configurations the FNE provides a model for statistical properties of force networks. Of course, in a real
physical system, the actual forces are selected by the history and
elasticity of the ladder or particles.

Strictly speaking, this indeterministic point of view only makes
sense in the limit of very hard contacts. A subtle point is then
that for perfectly rigid {\em frictionless} spheres,
the geometry completely specifies the forces --- frictionless hard
spheres in dimension $d$ organize such that their mean number of contacts $z$
reaches a well defined limit $z_{\rm iso} = 2d$, termed {\em
isostatic}, such that the number of contact forces and mechanical
constraints precisely balance \cite{alexander,moukarzel98,tkachenko99,roux00}. In more modern language,
the small relative deformations of the particles imply that
granular media are close to the ``jamming point''
\cite{liu98,ohern03,zhang05,vanhecke09}. Frictionless spheres are isostatic
in this limit. Frictional particles, however, are generally not
isostatic at the jamming point \cite{silbert02,kasahara04,unger05,somfai07}.
The idea of sampling all force configurations compatible with
force balance/torque balance and boundary conditions is therefore
on firmer footing physically in frictional packings. Having said
that, the ensemble is also mathematically well-defined for
frictionless systems, and many of the examples discussed below are
frictionless.

An additional motivation to study the FNE is somewhat more abstract. The idea of sampling all possible configurations goes back to Edwards, who advocated considering ensembles of all grain configurations consistent with some set of boundary conditions \cite{edwards89}. In general this is hard or impossible to do explicitly. The FNE can be seen as a restricted Edwards ensemble \cite{bouchaudleshouches}, with an extensive measure of the stress playing a role analogous to energy in equilibrium statistical mechanics \cite{snoeijer04b,tighe08b}. The ensemble then allows one to explore the consequences of and limits to the Edwards approach. For example, we will see that one popular notion, namely that entropy maximization implies robustly exponential force statistics \cite{edwards08}, does not survive confrontation with the FNE.

The scope of this paper is to review the literature on the FNE and
closely related work
~\cite{halsey99,snoeijer04a,snoeijer04b,mcnamara04,unger05,tighe05,ostojic05,snoeijer06,ostojic06,ostojic06b,ostojic07,vaneerd07,tighe08b,vaneerd09,shaebani09,tighe09b,tighe10a}. The outline
of this paper is as follows. We first motivate the ensemble in
more detail and quantify the degree of force indeterminacy in
Sec.~\ref{sec:motivation}. Statistical properties of the ensemble
are discussed in Sec.~\ref{sec:statistics}, with a focus on the
contact force distribution $P(f)$. Section~\ref{sec:mechanics} addresses the mechanical
response to applying an external load, such as a uniform shear
stress or a localized point force. The paper closes with a
discussion on the successes and limitations of the FNE, where we
also point out future directions. Throughout the paper we confront
the ensemble predictions to experimental or numerical data, when
available.

\section{Motivation: Force Indeterminacy}
\label{sec:motivation}

\begin{figure}[tbp]
\centering
\includegraphics[clip,width=0.85\linewidth]{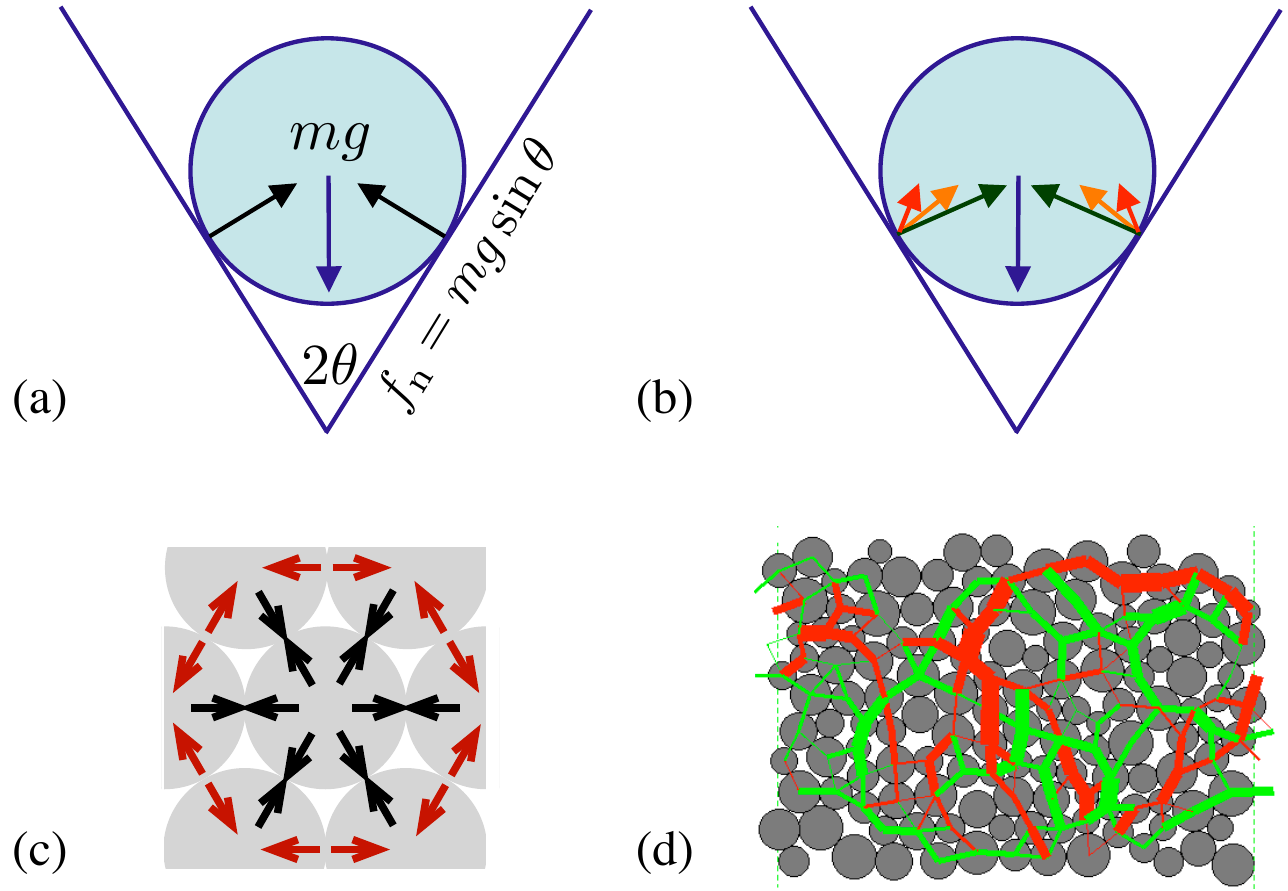}
\caption{(a) A rigid ball in a groove
\cite{halsey99,mcnamara04,vanhecke09} can be supported by purely
normal forces at the wall. (b) Introducing tangential, i.e~frictional, forces allows a range of balanced force configurations. (Adapted from \cite{vanhecke09}.) (c) Similar rearrangements, called ``wheel moves'' \cite{tighe05}, do not change the net vector force on a grain in the frictionless triangular lattice. (d) Rearrangements in disordered packings are delocalized. Here, changes to normal forces (color maps to sign, thickness to magnitude) of a rearrangement in a frictional packing; changes to tangential forces not shown. (From \cite{unger05}.)
 }
\label{fig:ur-wheelmoves}
\end{figure}

The FNE crucially relies on force indeterminacy, meaning that the
equations of local force and torque balance do not uniquely determine
the forces. It is therefore instructive to specify it further.
As an example, we consider a rigid ball in a groove with opening angle $2\theta$ and contact forces ${\bf f} =
(f_{\rm{n}}^{(1)}, f_{\rm{t}}^{(1)}, f_{\rm{n}}^{(2)},
f_{\rm{t}}^{(2)})$  (see Fig.~\ref{fig:ur-wheelmoves}). This system illustrates many salient features of indeterminacy~
\cite{halsey99,mcnamara04,vanhecke09}. By inspection, the weight
of the ball $-mg\hat{z}$ can be supported by purely normal forces
${\bf f}_0 = (c,0,c,0)$, where $c = \frac{1}{2}mg/\sin{\theta}$.
This solution is not unique, however; by adding normal and
tangential forces proportional to $\delta{\bf f} = (-\sin{\theta},
\cos{\theta},-\sin{\theta},\cos{\theta})$, other solutions of
the form ${\bf f} +\delta{\bf f} $ can be identified.

Indeterminacy carries over to packings of many grains, with the frictionless triangular lattice being an instructive example. There each grain participates in $z = 6$ contacts, so there are $z/2$ distinct contact forces per grain. As each grain brings $d=2$ force balance constraints, we anticipate $(z/2) - d = 1$ degree of freedom per grain, i.e.~one rearrangement of the forces that respects force balance, as with the ball in a wedge. This rearrangement, identified in Ref.~\cite{tighe05}, is called a wheel move (Fig.~\ref{fig:ur-wheelmoves}c). The idea is that by decreasing all six forces that contact a certain grain (black arrows) while simultaneously increasing the six forces that lie on a shell around the central grain by the same amount (red arrows), the total vector force on all grains remains invariant. Note that one should be careful that all contact forces remain noncohesive.

Wheel moves can be generalized for packings with disorder and friction \cite{unger05,tighe05,vaneerd07}. In this case the rearrangements of the forces generally cease to be localized, as demonstrated in  Fig.~\ref{fig:ur-wheelmoves}d. Just as in the frictionless triangular lattice, the number of independent force rearrangements $N_w$ in a packing of $N$ grains is given by the excess of force components over force/torque balance constraints,
\begin{equation}
N_w = \frac{1}{2}\Delta z \, d_f N + O(1).
\label{eqn:counting}
\end{equation}
Here $\Delta z = z - z_{\rm iso}$ and $d_f$ is the number of force components per contact. The isostatic contact number $z_{\rm iso} = 2d$ (resp.~$d+1$) and $d_f = 1$ (resp.~$d$) in  frictionless (frictional) sphere packings \cite{vanhecke09}. The correction in Eq.~(\ref{eqn:counting}) depends on details of the boundary conditions.

Refs.~\cite{tighe05} and \cite{vaneerd07} give a prescription for constructing a set of $N_w$ rearrangements $\lbrace \delta {\bf f}_k \rbrace$. With these in hand, any force network ${\bf f} = \lbrace {\vec f}_{ij} \rbrace$ on a given contact network can be expressed by giving the weight $w_k$ on each force rearrangement:
\begin{equation}
{\bf f} = {\bf f}_0 + \sum_{k=1}^{N_w} w_k\,\delta{\bf f}_k \,.
\label{eqn:fnetwork}
\end{equation}
${\bf f}_0$ is any balanced force network with the appropriate stress tensor $\hat \sigma$, i.e.~a particular solution. The stress tensor can be expressed in terms of the microscopic forces via \cite{snoeijer04a}
\begin{equation}
\sigma_{\alpha \beta} = \frac{1}{2V} \sum_{ij} f_{ij,\alpha} r_{ij,\beta} \,.
\end{equation}
Here $V$ is the volume of the contact network and ${\vec r}_{ij}$ points from the center of grain $i$ to the center of grain $j$. 

The force rearrangements $\lbrace \delta {\bf f}_k \rbrace$ alter forces in the system without violating mechanical equilibrium, and are therefore the mechanism of force fluctuations in the FNE. They can be employed as Monte Carlo moves to sample the space of force networks \cite{tighe05,vaneerd09}, which can be achieved by varying the $\lbrace w_k \rbrace$. It is important to note that the weights $\lbrace w_{k} \rbrace$ are strongly constrained by inequalities. These express the constraints that ($i$) all normal forces are noncohesive, i.e.~$f_{\rm n} \ge 0$, and ($ii$) all tangential forces respect Coulomb's constraint $|f_{\rm t}| \le \mu f_{\rm n}$. 

As the FNE employs a flat measure, meaning all valid force networks are given equal statistical weight, all the essential physics of the FNE is encoded in the geometry of a high-dimensional space. The geometry, in turn, is determined by force balance and the positivity and Coulomb constraints. One can think of each force network as occupying a point in a space with each weight $w_k$ describing an axis. In this space the noncohesive and Coulomb constraints each describe planar boundaries where a normal force is zero or a tangential force is fully mobilized, respectively. These boundaries enclose a convex polytope, which we call ${\cal F}_C$ and which contains all possible force networks $\lbrace {\bf f}_k \rbrace$ on a contact network $C$ \cite{mcnamara04,unger05,snoeijer04b}.

Indeterminacy is intimately connected with the size and shape of
the space of force networks ${\cal F}_C$. Globally, this space has
dimension $D=N_w$. Locally, the ``size'' of the space in a
particular direction can be quantified by seeking the largest and
smallest values a particular contact force component can take on, by
varying the $\lbrace w_k \rbrace$ to the extremes of ${\cal F}_C$ \cite{mcnamara04,shaebani09}.
E.g.~for the tangential force at contact $c$, the indeterminacy $\eta$
is
\begin{equation}
\eta(f_{\rm t}^c) = \frac{ (f_{\rm t}^c)^{\rm max} - (f_{\rm t}^c)^{\rm min} }
{\frac{1}{2} \left[ (f_{\rm t}^c)^{\rm max} + (f_{\rm t}^c)^{\rm min} \right] } \,,
\label{eqn:indeterminacy}
\end{equation}
The denominator serves simply to normalize by a typical force scale in the packing, hence other choices are possible. McNamara and Herrmann, who study packings under gravity using Contact Dynamics (CD) and MD, use the average weight of a grain  $\langle m g \rangle$ \cite{mcnamara04}. It is apparent from Fig.~\ref{fig:ur-wheelmoves}b that force fluctuations have a rich spatial structure, and indeed Ref.~\cite{mcnamara04} finds that $\eta$ is broadly distributed and is positively correlated with contacts that carry large forces. Note that, although there is an upper bound on the maximum normal force $f_{\rm n}$ (and hence  $|f_{\rm t}|$) in a finite size packing \cite{tighe05,vaneerd09}, that bound grows with system size. It may therefore be useful to study the behavior of $\eta$ in the thermodynamic limit.

\begin{figure}[tbp]
\centering
\includegraphics[clip,width=1.0\linewidth]{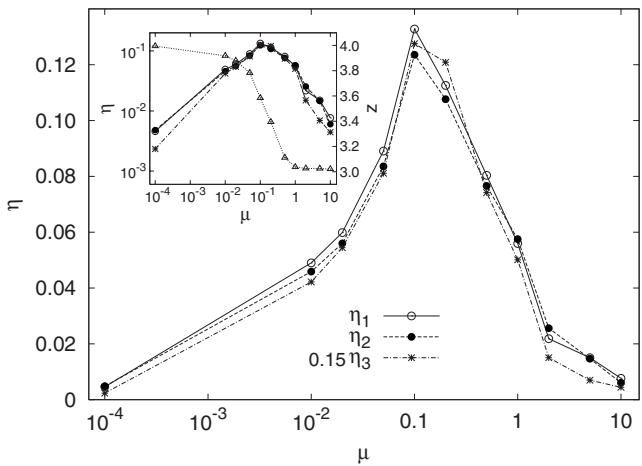}
\caption{
Dependence of the force indeterminacy $\eta$ on friction coefficient $\mu$ for contact networks equilibrated under Contact Dynamics \cite{shaebani09}. Several different measures of force indeterminacy show the same qualitative evolution with $\mu$.  $\eta_1$ is related to the relative fluctuations of contact force components when valid networks are randomly sampled, while $\eta_2$ depends on the mean Euclidean distance between randomly sampled networks. See Ref.~\cite{shaebani09} for precise definitions. $\eta_3$ is given by Eq.~(\ref{eqn:indeterminacy}). (inset) The peak in $\eta$
corresponds with a sharp decrease in mean contact number $z$. (From \cite{shaebani09}.)
} 
\label{fig:indeterminacy}
\end{figure}

Averaging $\eta$ of Eq.~(\ref{eqn:indeterminacy}) over all normal and tangential contact force components gives a purely geometric global measure of indeterminacy, related to the volume of ${\cal F}_C$. As shown in Fig.~\ref{fig:indeterminacy}, several alternative measures display qualitatively similar, nontrivial dependence on friction coefficient $\mu$ in frictional packings generated by CD, which treats perfectly rigid grains \cite{unger05,shaebani09}. Indeterminacy must be zero when the system is isostatic, which generically occurs~for $\mu = 0$ and $\mu \rightarrow \infty$ \cite{vanhecke09}. For finite $\mu$ the global indeterminacy $\langle \eta \rangle$ displays a maximum for $\mu \approx 0.1$ due to a balance between two competing effects \cite{unger05}. Increasing $\mu$ opens the Coulomb cone, which increases the volume of ${\cal F}_C$ without changing its dimension. In contrast, increasing $\mu$ lowers the contact number $z$, and hence $D$. For a number of numerical protocols $z(\mu)$ decreases abruptly around $\mu \approx 0.1$. Thus the force indeterminacy $\langle \eta \rangle$ displays a sharp signature of the packing structure (Fig.~\ref{fig:indeterminacy}).

\section{Stress statistics}\label{sec:statistics}

The statistics of local measures of the stress, such as the force $f$ at a contact or the pressure $p$ on a grain, provide a fundamental microscopic characterization of the material's stress state. The FNE has turned out to be an ideal model to probe these probability distributions.

We first provide a brief review of characteristic properties of the stress statistics of static packings. The second part of this section addresses theoretical aspects of statistics in the FNE, followed by a discussion.

\subsection{Characteristic features of $\pf$}

\begin{figure}[tbp]
\centering
\includegraphics[clip,width=1.0\linewidth]{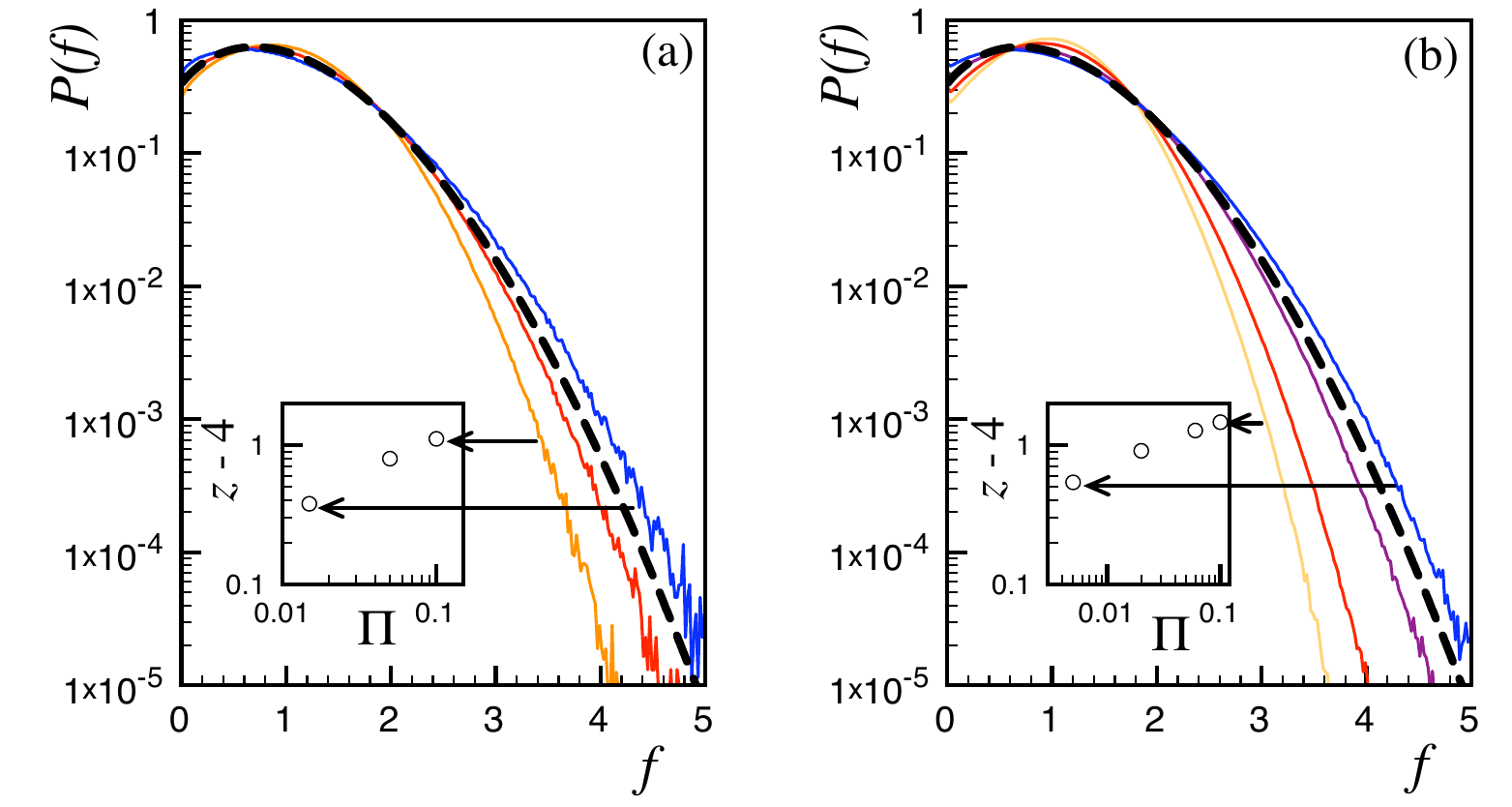}
\caption{$\pf$ for Molecular Dynamics simulations
of soft spheres in 2D (solid curves) and the FNE on the frictionless triangular
lattice (dashed curves). $f$ is normalized to 
$\langle f \rangle$. MD grains interact via one-sided (a) Hookean or (b) Hertzian springs. (insets) MD
distributions differ in their confining pressure $\Pi$, which
tunes the contact number $z$ relative to the isostatic value
$z_{\rm iso} = 4$. $\pf$ in the FNE shows strong qualitative similarity to $\pf$ in MD for both force laws. (Adapted from \cite{vaneerd09}.)
} 
\label{fig:mdPf}
\end{figure}

Let us begin with a numerical characterization of the force
probability density $\pf$ from Molecular Dynamics (MD) simulations on
frictionless systems. Figure~\ref{fig:mdPf} depicts $\pf$ for
a range of confining pressures $\Pi$, effectively changing the
coordination number, for both Hookean and Hertzian interactions \cite{vaneerd09}. Several features stand out. All distributions have ($i$) a peak near the mean force $\langle f \rangle$, ($ii$) a nonvanishing
weight as $f \rightarrow 0^+$, and ($iii$) a width comparable to
$\langle f \rangle$. The latter two properties reflect
heterogeneity in the force network, while it has  been suggested
that the peak is symptomatic of the jammed state \cite{ohern01}. These three features, the
``look and feel'' of $\pf$, are observed in a variety of
simulations and experiments
\cite{radjai96,makse00,tkachenko00,ohern01,silbert02,gg04,liu95,mueth98,tsoungui98,lovoll99,brujic03,majmudar05,zhou06}. This hints at generic mechanisms that can
be probed within the FNE, in particular since the qualitative
features of $\pf$ in packings are insensitive to the force law.

The simplest system to which one can apply the FNE is the
two-dimensional contact network with triangular lattice symmetry
and isotropic stress~\cite{snoeijer04a,tighe05,vaneerd07}, as in
Fig.~\ref{fig:FNE}a. The corresponding $\pf$ is represented by the
dashed curves in Fig.~\ref{fig:mdPf}. Indeed, the FNE prediction
agrees very well with the results from Molecular Dynamics
simulations and bears the three characteristic features of $\pf$.
In addition, very similar results were obtained when numerically
sampling FNE for disordered disk packings with varying average
contact numbers ranging from $z = 6$ to 4.3~\cite{vaneerd07}.
Closer approach to the isostatic point $z_{\rm iso}=4$, where the
space of allowed force networks vanishes, is numerically
impractical; hence the FNE is not the appropriate tool to probe
$\pf$ near $z_{\rm iso}$.

\begin{figure}[tbp]
\centering
\includegraphics[clip,width=0.8\linewidth]{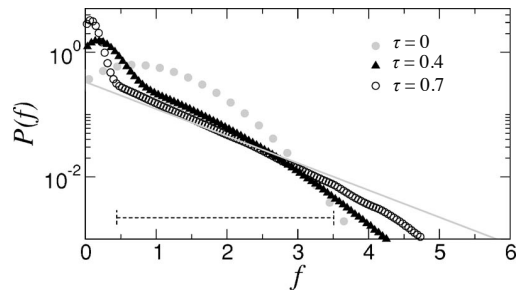}
\caption{$\pf$ in the frictionless triangular lattice subject to a shear stress displays a quasi-exponential intermediate regime that grows with increasing anisotropy $\tau$. The distribution approaches a pure exponential (gray curve) in the asymptotic limit $\tau \rightarrow 1$. (Adapted from \cite{tighe05}.)  }
\label{fig:aniosoPf}
\end{figure}

Ref.~\cite{tighe05} explores the consequences of stress anisotropy
in the triangular lattice. Anisotropy corresponds to $\tau > 0$
when the principle stresses $\sigma_1 \ge \sigma_2$ are unequal:
\begin{equation}
\tau= \frac{\sigma_1 - \sigma_2}{\sigma_1 + \sigma_2} \,.
\label{eqn:aniso}
\end{equation}
Anisotropy introduces a separate force distribution for each lattice direction. $\pf$ for forces aligned with the principal stress direction broadens and develops a regime of exponential decay before crossing over to faster-than-exponential decay (Fig.~\ref{fig:aniosoPf}). The width of this regime grows with increasing $\tau$. For sufficiently strong anisotropy, $\pf$ averaged over lattice directions loses its peak near $\langle f \rangle$, suggesting the presence of a peak in $\pf$ is not a robust signature of the jammed state \cite{snoeijer04b,tighe05}. In an ordered packing in the limit $\tau \rightarrow 1$, force is restricted to one lattice direction, the system is quasi-one dimensional, and $\pf$ becomes exactly exponential \cite{tighe05}. A similar exponential regime is observed for disordered contact networks under shear, before crossing over to asymptotically Gaussian decay \cite{vaneerd07}.

\begin{figure}[tbp]
\centering
\includegraphics[clip,width=0.98\linewidth]{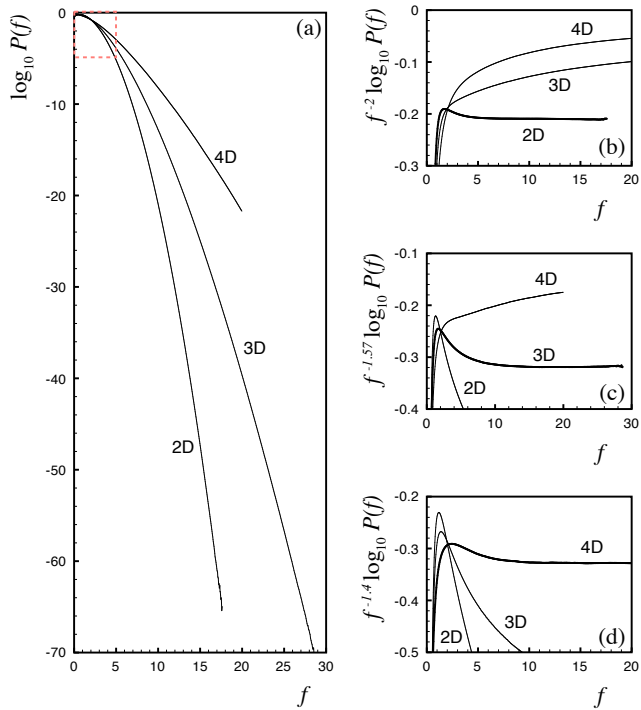}
\caption{$\pf$ from umbrella sampling in the frictionless triangular lattice (2D), frictionless fcc lattice (3D), and a disordered packing in four dimensions (4D) with contact number $z \approx 20.9$. The dashed box indicates the plotting range in Fig.~\ref{fig:mdPf}. (a) $\log{\pf}$ bends downward when plotted versus $f$, indicating faster than exponential decay. 
If $\pf \sim \exp{(-f^b)}$ then $f^{-b}\log{\pf}$ approaches a constant for large $f$. We observe (b) $b = 2.0(1)$ (Gaussian decay) in 2D, (c) $b = 1.6(1)$ in 3D, and (d) $b = 1.4(1)$ in 4D. (Adapted from \cite{vaneerd07}.)
} 
\label{fig:Pfdim}
\end{figure}

Recently, numerical results for the FNE have been obtained using umbrella sampling, with which the FNE can be sampled far more efficiently --- and hence $\pf$ can be determined far more accurately --- than ensembles generated by experiments or MD \cite{vaneerd07,tighe08b,vaneerd09}. With this accuracy, which permits sampling $\pf$ over tens of decades in the tail (Fig.~\ref{fig:Pfdim}a), it has been put beyond any doubt that $\pf$ in the FNE decays faster than exponentially. This is already suggested by Figs.~\ref{fig:mdPf}, in which $\pf$ clearly bends downwards on a semi-log plot. $\pf$ in the FNE is numerically determined in Ref.~\cite{vaneerd07} to decay as $\exp{(-f^b)}$, with the exponent $b$ dependent on the dimension. Consistent with theoretical arguments (see below), $b=2$ for two dimensions (Fig.~\ref{fig:Pfdim}b), i.e.~the tail of $\pf$ is Gaussian. For higher dimensions $\pf$ decays as a compressed exponential (Fig.~\ref{fig:Pfdim}c,d), with $b \approx 1.6(1)$ in 3D and $b \approx 1.4(1)$ in 4D. Numerics show no dependence of $b$ on disorder for $\Delta z$ as small as 0.3 \cite{vaneerd07,vaneerd09}. In light of early experiments finding exponential tails, these results are surprising; we return to this point below.

To summarize, $P(0)>0$ for all frictionless force networks in the FNE. Though there is often a peak for finite $f$, it can be destroyed by strong anisotropy. The force distribution is always wide and its asymptotic tail is set only by the dimension. Strong anisotropy, however, can open an intermediate regime of exponential decay.

\subsection{$\pf$ in theory}

Two of the robust features of $\pf$ identified above are the
finite value of $\pf$ for vanishing force, and the asymptotic form
of its tail. We will discuss now how, in the FNE, the origins of the former remain an open quesion, while the latter follows from entropy maximization
in the presence of an unanticipated constraint.

It has been suggested that, in the presence of a flat measure, the finite value of $P(f)$ for vanishing force can be traced to the geometric properties of the high dimensional space  ${\cal F}_C$ \cite{snoeijer04b}. 
In frictionless systems, states with a zero contact force sit on one of the boundary facets of ${\cal F}_C$, and high dimensional spaces have the curious but well known property that the overwhelming majority of the space is close to the boundary. 
This fact seems to motivate the presence of many vanishingly small forces. In preparing this review, however, it became apparent that this reasoning does not stand up to closer scrutiny. Simply by dimensional analysis, the fraction $v(\delta f)$ of the volume of ${\cal F}_C$ within $\delta f$ of the surface  must be $v \simeq \frac{\Delta z}{z} \frac{\delta f}{\langle f \rangle}$. If this volume were simply divided  among the $N_{\rm c}$ contact forces, then $P(0)  \approx  v(\delta f)/N_{\rm c} \delta f$ would vanish in the thermodynamic limit. The fact that it does not means that  a typical force network within $\delta f$ of one boundary facet is also within $\delta f$ of a finite fraction of the other boundary facets. Precisely how this can be understood, and how, e.g., $P(0)$ depends on contact number $z$, is an interesting and open question.

\begin{figure}[tbp]
\centering
\includegraphics[clip,width=0.9\linewidth]{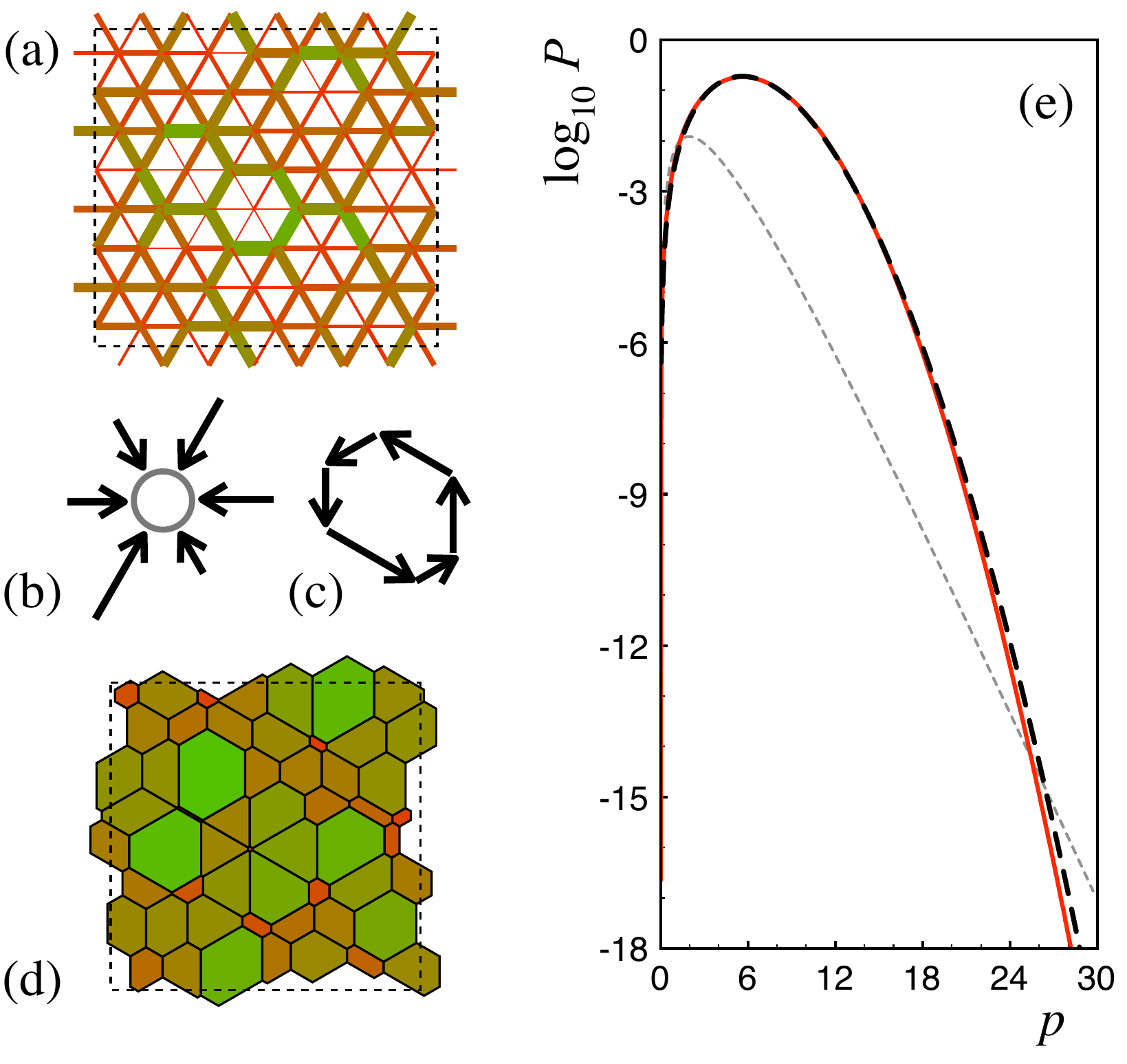}
\caption{(a) Periodic force network on the frictionless triangular lattice. (b) Forces on each grain construct a tile (c) by summing them graphically. Tiles tessellate space due to Newton's laws.  (d) The reciprocal tiling of (a). (From Ref.~\cite{tighe10a}.) (e) The numerical distribution $P(p)$ of the pressure on a grain in the triangular lattice (solid curve) is well described by Eq.~(\ref{eqn:prediction}) (black dashed curve) but not by a comparable calculation neglecting conservation of tiling area (gray dashed curve). (Adapted from \cite{tighe08b}.) }
\label{fig:tiling}
\end{figure}

We have seen that $\pf$ in two dimensions has a Gaussian tail (Fig.~\ref{fig:Pfdim}b). To understand how this comes about, it is instructive to study a 
related distribution, namely that of local pressures $p$.
The grain scale ``pressures'' $p_i = \sum_j (f_{\rm n})_{ij}$ are convenient both because the constraints of force
balance enter at the grain scale and because the pressure is a slightly
coarse-grained stress.
The form of the probability distribution $P(p)$ is motivated in Ref.~\cite{tighe08b} by an entropy maximization argument,
the crux of which we sketch here. 
The argument relies on the observation that force networks in two dimensions admit a reciprocal representation in which
forces from the network are used to construct tiles that
tessellate space \cite{maxwell1864}. A tile is formed by graphically summing the
contact forces on a grain in a right-hand fashion;
Fig.~\ref{fig:tiling} gives an example. The key observation is that, because stress $\hat \sigma$ and volume $V$ are both fixed in the FNE, the area of the tiling  ${\cal A} = ({\rm det}\,{\hat \sigma})V$ is {\em also} invariant \cite{tighe08b,tighe09b,tighe10a}. This is a collective effect; the tiling only exists when every grain is in force balance. 

For concreteness, consider a frictionless triangular lattice with isotropic stress and pressure, ${\hat \sigma} := (\sqrt{3}/6r)\Pi\, \mathbbm{1}$, where $r$ is the grain radius. Because the stress is fixed, the grain-scale pressures $\lbrace p_i \rbrace$ obey a sum rule:
\begin{equation}
\sum_{i} p_{i} = \Pi N  \,.
\label{eqn:sumpressure}
\end{equation}
$\Pi$ is therefore the average local pressure.
The new observation is that because there is a tiling and its area is invariant, the local areas $\lbrace a_i \rbrace$ {\em also} obey a sum rule
\begin{equation}
\sum_i a_i = \sqrt{3}\left(\frac{\Pi}{6}\right)^2 N\, . 
\label{eqn:sumarea}
\end{equation}
Note that Eq.~(\ref{eqn:sumarea}) is satisfied automatically when Eq.~(\ref{eqn:sumpressure}) {\em and} force balance on every grain are imposed. In analytical calculations one typically resorts to treating local force balance approximately, with the consequence that Eq.~(\ref{eqn:sumarea}) is violated -- see Ref.~\cite{tighe10a} for a detailed discussion of this point. In this case, imposing the area sum rule as a constraint reintroduces a necessary consequence of local force balance. The surprise is that by incorporating this global constraint, instead of the ${\cal O}(N)$ local force balance constraints, one can successfully predict stress statistics in the FNE via an entropy maximization calculation \cite{tighe08b}.

The resulting distribution of local pressures in the frictionless triangular lattice is
\begin{equation}
P(p) = Z^{-1} p^\nu e^{-\alpha p -  \gamma \langle a(p) \rangle }\,.
\label{eqn:prediction}
\end{equation}
The exponent in the prefactor depends on contact network; $\nu = 3$ for the for the frictionless triangular lattice.
$Z$, $\alpha$, and $\gamma$ are not free parameters but Lagrange multipliers determined by 
normalization of $P(p)$ and Eqs.~(\ref{eqn:sumpressure}) and
(\ref{eqn:sumarea}). $\langle a(p) \rangle \propto p^2$ is the
average area of a tile with corresponding pressure $p$; it appears because the area sum rule has been imposed. For asymptotically large pressures the quadratic term dominates and the tail of $P(p)$ is Gaussian. For small $\gamma$, however, the Gaussian form may only become apparent deep in the tail. Eq.~(\ref{eqn:prediction}) is in excellent
agreement with numerics; see Fig.~\ref{fig:tiling}d. Though it has not been shown that a Gaussian tail in $P(p)$ requires a Gaussian tail in $\pf$, it is an empirical fact that the tails of $\pf$ and $P(p)$ always have the same form. Hence the invariant tiling area presumably also explains the tail of $\pf$.

The pressure sum rule, Eq.~(\ref{eqn:sumpressure}), is reminiscent of energy in a microcanonical ensemble, which also obeys a sum rule. This similarity has provoked a number of authors to explore parallels between equilibrium statistical mechanics and stress-based ensembles of packings, under the assumption that they are also entropy maximizing, see e.g.~Refs.~\cite{evesque99,kruyt02,bagi03,ngan03,goddard04,henkes07,metzger08}. In a statistical mechanics framework, a natural first step is to consider the analog of an ideal gas, i.e.~completely neglecting correlations between particles and maximizing entropy in the presence of the constraint Eq.~(\ref{eqn:sumpressure}). This ``ideal gas approach'' predicts exponential tails, and is therefore not adequate to describe statistics of forces in the FNE.  Though the calculation of Ref.~\cite{tighe08b} sketched above also neglects spatial correlations, by incorporating the tiling area constraint it retains information that the ideal gas approach throws out. It is remarkable that including just one more global constraint, rather than the many local force balance constraints, so thoroughly captures the numerical distribution (Fig.~\ref{fig:tiling}d). Neglecting spatial correlations is no longer reasonable in the presence of a diverging length scale near isostaticity \cite{wyart05,ellenbroek06}, so Eq.~(\ref{eqn:prediction}) makes no prediction for this case. 

Is there a counterpart to the tiling constraint for higher
dimensions? One can indeed construct polyhedra analogous to the
polygonal tiles of Fig.~\ref{fig:tiling}
\cite{minkowski,tighe08b}, but whether the sum of their
volumes is conserved remains open. If so, one would have $P(f)
\sim \exp{(-f^b)}$ with $b = d/(d-1)$. Encouragingly, this is in
reasonable agreement with the results of Fig.~\ref{fig:Pfdim}.

\subsection{Comparison to experiment and simulation}
How well do these features conform with $\pf$ from experimental or
simulated systems? Finite $P(0)$ is indeed a robust feature of
bulk measurements. Similarly, sampled distributions are broad, and
a qualitative change in $\pf$ has been observed in the only
experiments to systematically vary anisotropy \cite{majmudar05}.
It is less clear if any general statements can be made about the
form of the tail of $\pf$ in real granular systems \cite{radjai96,radjai99,makse00,tkachenko00,ohern01,silbert02,ohern02,gg04,zhang05,vaneerd09,liu95,mueth98,tsoungui98,lovoll99,blair01,erikson02,brujic03,majmudar05,zhou06}.

The earliest measurements of $\pf$ --- made at the boundary of packings using the carbon paper method \cite{liu95,mueth98} --- displayed unambiguously exponential tails. Accompanied by the q-model \cite{coppersmith96}, which predicts exponentials, these experiments established the expectation that $\pf$ should decay exponentially. Forces in the bulk are difficult to access experimentally, and there have been few measurements \cite{tsoungui98,brujic03,majmudar05,zhou06}. Those there are raise doubts regarding the robustness of exponentials; distributions from isotropic 2D photoelastic systems \cite{majmudar05} and 3D emulsions \cite{brujic03,zhou06} show downward curvature on a semi-log plot, suggesting faster than exponential decay. Statistics are limited, however, and it is difficult to determine the asymptotic form of the tail.

Simulation results are inconsistent \cite{radjai96,radjai99,makse00,tkachenko00,ohern01,silbert02,ohern02,gg04,zhang05,silbert06,vaneerd09}. Though some distributions are clearly
exponential \cite{radjai96,radjai99}, others display noticeable curvature on a semi-log plot \cite{ohern02,gg04,silbert06,vaneerd09}. Few simulations capture more than three decades in the tail, making it difficult to distinguish exponentials from broad Gaussians or compressed exponentials. The tail of $\pf$ may show signatures of the approach to isostaticity; this point is not settled, and the FNE, which vanishes at the isostatic point, provides no illumination. While data from Zhang and Makse cross over from Gaussian to exponential decay near unjamming \cite{zhang05}, data from Silbert et al.~remain Gaussian even when the distance to the critical packing fraction $\phi_c$ is as small as $10^{-6}$ \cite{silbert06}. O'Hern et al.~find Gaussian tails in an ensemble at fixed distance to the transition, while fluctuations in this distance due to finite size, which occur in fixed volume ensembles, can render tails exponential \cite{ohern02}.

\section{Mechanical properties}\label{sec:mechanics}

It is generally thought that force networks are
important for the mechanical properties of static and quasistatic
granular materials. Since the FNE accurately describes the
statistics of real force networks, one could ask whether it also
captures mechanical response to external loading. This can indeed
be explored within the FNE. We discuss the response to
anisotropic loads (shear stress) and localized loads (a point force at the boundary).

\subsection{Response to a shear stress}

\begin{figure}[tbp]
\centering
\includegraphics[clip,width=0.8\linewidth]{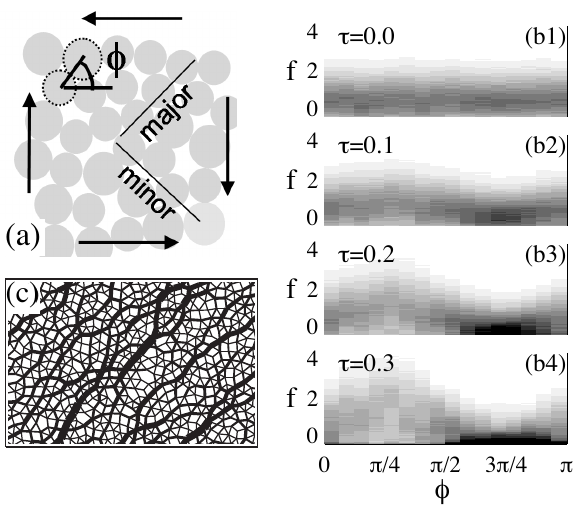}
\caption{(a) Shear stress breaks isotropy, introducing a major and minor principal stress axis. (b) Density plot of the angle-dependent force distribution $P_\phi(f)$ for increasing anisotropy $\tau$. (c) A strongly anisotropic force network. Adapted from Ref.~\cite{snoeijer06}.}
\label{fig:illustration}
\end{figure}

Anisotropic force states such as shown in Fig.~\ref{fig:aniosoPf} appear naturally when imposing a shear stress to the system. It is interesting to study this effect for disordered packings, for which the anisotropy cannot be aligned along preferred lattice directions. Figure~\ref{fig:illustration}a illustrates how the imposed shear stress $\sigma_{xy}$ gives rise to major and minor principal axes at angles $\phi = \pi/4$ and $3\pi/4$ respectively \cite{snoeijer06}. As above, we quantify anisotropy by $\tau$ from Eq.~\ref{eqn:aniso}.

The resulting force anisotropy can be investigated using the angle resolved PDF, representing $\pf$ at different orientations $\phi$. Figure~\ref{fig:illustration}b shows results obtained in the FNE for frictionless packings with isotropic contact networks \cite{snoeijer06}. The modulation along $\phi$ increases with the imposed value of $\tau$, signaling the increase of force anisotropy. This is further quantified through the angle resolved force average, $\bar{f} (\phi)$, which can be expanded in a Fourier series as in Refs.~\cite{radjai98,snoeijer06}.
\begin{equation}\label{barphi}
\bar{f}(\phi) = 1 + 2\tau \sin 2\phi - b_2 \cos 4 \phi +  \cdots.
\end{equation}
The second order coefficient is directly proportional to $\tau$, while the higher order terms do not
couple to the stress at all. Indeed, Eq.~(\ref{barphi})
already provides an accurate fit to the numerical results when
truncated at fourth order.

\begin{figure}[tbp]
\centering
\includegraphics[clip,width=0.5\linewidth]{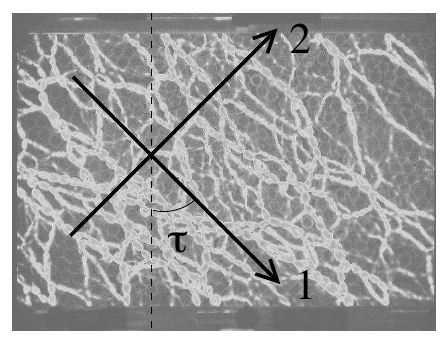}
\caption{Photelastic granular packing under shear. The strongest forces align with the major principal axis. (From Ref.~\cite{atman05}.)}
\label{fig:ericbob}
\end{figure}

Force anisotropy has been observed
experimentally~\cite{atman05,majmudar05} and in contact dynamics
simulations~\cite{radjai98}. Figure~\ref{fig:ericbob} shows an
experiment on sheared packings of photoelasic grains, which
visualizes the forces in the system~\cite{atman05}. These indeed
reveal the preferred orientation of large forces along $45^\circ$
with respect to the horizontal, coinciding with the major
principal axis.

Another striking feature of Fig.~\ref{fig:illustration}b is that
many contact forces along the minor axis  evolve towards zero for
increasing shear (black area in Fig.~7b3-b4 near $\phi \sim
3\pi/4$). These small forces are close to breaking (they cannot
become tensile), which will eventually lead to failure of the
system. Thus there must be a maximum $\tau_m$ above which no solutions exist. 
One can interpret this in terms of the volume spanned by all force
configurations, which continuously shrinks with $\tau$ and reaches
zero at $\tau_m$. It is found numerically that $\tau_m$ strongly
depends on the coordination number $z$, and approximately follows
the scaling~\cite{snoeijer06}
\begin{equation}
\tau _m \approx 2 \frac{z-z_{\rm iso}}{z},
\label{eqn:yieldstress}
\end{equation}
where $z_{\rm iso}$ is the isostatic coordination number. One thus finds that the maximum possible stress vanishes when the packing approaches the isostatic limit. This is strongly reminiscent of the onset of bulk modulus, shear modulus and dynamic yield stress at the jamming transition. 

Far away from the isostatic point, it is possible to apply mean field arguments to estimate the maximum shear stress in the FNE~\cite{ellenbroek07}. This is done by requiring the average force  $\bar{f}(\phi)$ to be positive in all directions $\phi$. This condition is of course much weaker than the requirement that all individual forces be positive, and therefore leads to an upper bound for $\tau_m$. Generalizing the argument to frictional particles with friction coefficient $\mu$, one estimates~\cite{ellenbroek07}

\begin{equation}
\tau_m < \frac{1+\sqrt{1+3\mu^2} }{3}.
\end{equation}
Simulations of the FNE with $z=5.5$ show that this upper bound is approached to within a few percent for $\mu = 0.5$ and $1.0$~\cite{vaneerd09}.

\subsection{Response to a point force}

Another way to assess the mechanical behavior is through the
response to a localized force. For small enough forces this probes
the Green's function of the granular packing and provides crucial
information on the effective continuum description of the system.
Experiments \cite{geng01,serero01,mueggenburg02} and simulations \cite{gg02,kasahara04,gland05} have shown that the spreading of the
load inside the material is not universal but can be along a
single broad peak, as is the case for isotropic linear elastic materials, or
more anisotropically along two peaks. The response is found to depend on
parameters such as friction coefficient, degree of disorder,
coordination number and amplitude of applied
force.

\begin{figure}[tbp]
\centering
\includegraphics[clip,width=0.99\linewidth]{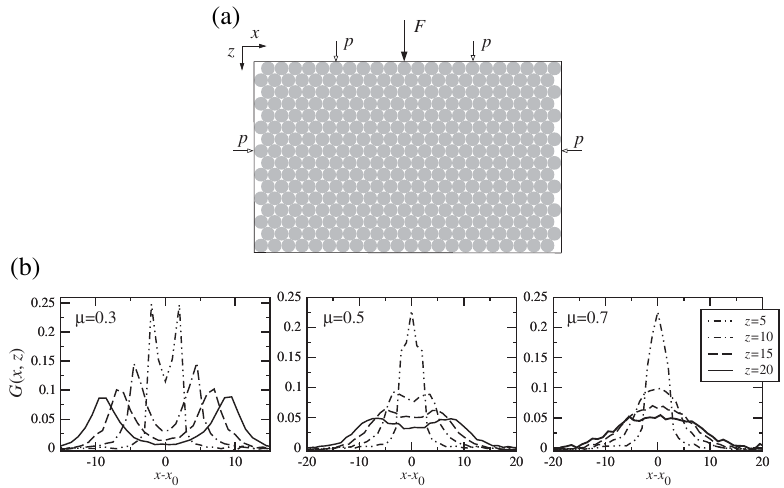}
\caption{(a) Contact network and boundary conditions for point force response. An otherwise isotropic system is augmented by a normal force $F$ at one boundary.  (b) Evolution  of the response with depth $z$ changes qualitatively with microscopic friction coefficient $\mu$. (Adapted from \cite{ostojic06}.)}
\label{fig:response}
\end{figure}

The response problem was addressed in the FNE for two-dimensional lattices with a free top surface~\cite{ostojic05,ostojic06}. Frictionless grains were studied on a triangular lattice, while a square lattice was used for frictional systems. Here we discuss the findings of~\cite{ostojic06}, where the packing was first put under isotropic pressure $p$ before applying a load $f$ on a grain in the top layer (Fig.~\ref{fig:response}). Within this setting it is possible to investigate the effect of the relative force amplitude, $F = f/p$, and the friction coefficient $\mu$. The response function $G(x,z)$ is defined by the difference in vertical force $W(x,z)$ on a grain with and without the point force:
\begin{equation}
G(x,z) = \frac{\langle W_{x,z}\rangle_F - \langle W_{x,z}\rangle_0}{F}.
\end{equation}
The brackets with subscripts $F$ and $0$ denote the average in the corresponding FNE.

A linear regime was found for small enough load, $F \lesssim 3$, where the response is independent of the amplitude $F$~\cite{ostojic06}. Figure~\ref{fig:response} displays $G(x,z)$ on the square lattice for various friction coefficients $\mu$, all with $F=3$. Each graph contains the response at different distances to the top surface and reveals how the load spreads through the material. Clearly, the response evolves from `two peaks' to `one peak' upon increasing $\mu$. This can be interpreted as follows. In the particular case where $\mu=0$, the response must be along the two downward lattice directions. For $\mu \neq 0$, the presence of tangential forces makes it possible to spread the load. This `freedom' increases with $\mu$ and eventually yields a single peaked response, as in isotropic elastic media. These observations agree well with MD simulations of ordered layers of grains~\cite{gg02,gg05}, where the grain-grain interactions were modeled in detail. These simulations showed that friction strongly enhances the regime where the material behaves like a linear elastic solid.

It is interesting to compare the $\mu=0$ results for the square lattice ($z=4$) and the triangular lattice ($z=6$). Though it is dangerous to extrapolate generic properties of isostatic systems from lattice packings, it is noteworthy that the response transitions from single-peaked (not shown) to double-peaked as $(z -  z_{\rm iso}) \rightarrow 0^+$. This is consistent with the emerging picture that there is a diverging length scale $\ell^\star \sim 1/\Delta z$ above which a packing can be viewed as an elastic continuum \cite{ellenbroek06}. Anisotropy may also play a role. The triangular lattice is isotropic in linear elasticity while the square lattice is not, and anisotropic continua admit two-peaked response  \cite{otto03,tighe08a}.
Finally, at fixed depth the response gradually changes to two peaks when increasing $F$ well into the nonlinear regime, also consistent with continuum descriptions \cite{tighe08a}. This crossover occurs when the local load $W$ becomes much larger than the horizontal pressure scale, in which case the horizontal forces hardly contribute the force balance. Locally, this effectively changes the lattice from `triangular' to `square', the latter allowing only for transmission along the lattice directions -- see also Ref.~\cite{gg02}.

\section{Conclusions and Outlook}

The FNE is a convenient minimal model system for static granular materials that takes into account local force balance explicitly. Due to its simplicity its properties can be computed accurately.

The FNE reproduces the most robust features of the statistics of contact forces, suggesting that these features follow from the geometry of a high-dimensional space, and that details of the force law are of secondary importance. Perhaps most surprising is the finding that $\pf$ in the FNE decays faster than exponentially. It is our impression that consensus has crystallized about the notion that exponential tails are a hallmark of granular force fluctuations, motivated by early experimental \cite{liu95,mueth98}, numerical \cite{radjai96}, and theoretical \cite{coppersmith96} work. Although there is support for this view, it seems to outrun the available evidence. The FNE suggests an alternative perspective; namely that distributions decay faster than exponentially asymptotically but, due to anisotropy, may appear exponential over accessible ranges. One useful role for the FNE is that of litmus test: a model that cannot explain results in the FNE is too simplistic. On this basis the q-model and ideal gas-like extrapolations from the Edwards ensemble can already be rejected. Therefore, {\em if} the tail of $\pf$ in real systems is robustly exponential, a theoretical explanation is still lacking. 

As a statistical measure, $\pf$ carries no information about the spatial structure of force networks. Studies of thresholded force networks offered the intriguing suggestion that systems with vector force balance represent a different universality class from ordinary percolation \cite{ostojic06b,ostojic07}. More recent work, however, attributes these observations to crossover effects \cite{bernard}. It is therefore an open question whether force networks carry diverging spatial signatures of the impending loss of rigidity as the isostatic point is approached, as do the vibrational \cite{wyart05} and response \cite{ellenbroek06} properties of soft sphere packings.

Studies of response hint at a connection between hyperstaticity of forces and elasticity or continuum-like response in the corresponding soft sphere packing \cite{ellenbroek06}. It remains an interesting and open question whether the stress statistics of Eq.~\ref{eqn:prediction} break down near isostaticity.

Important topics for future studies are the nature of force networks in frictional systems, in systems of non-spherical particles, and in flowing systems --- does the FNE also capture the statistics of these systems? Finally, one may wonder if the idea of flatly sampling a family of configurations, which the FNE does for force configurations, can be extended to other cases, such as the family of contact geometries that correspond to a certain contact topology.

\section*{Acknowledgments}
BPT acknowledges support from the Dutch physics foundation FOM.

\bibliographystyle{apsrev}
\bibliography{granular.bib}

\end{document}